\documentclass{PoS}
\usepackage{axodraw}
\usepackage{cite}
\usepackage{axodraw}

\newcommand{\propover}{\raise0.5ex\hbox{\kern0.8em$\large\propto$\kern-1.35em\raise-1.5ex\hbox{\tiny{$p^2\to 0$}}}}
\newcommand{\toover}{\raise0.5ex\hbox{\kern0.8em$\large\to$\kern-1.35em\raise-1.5ex\hbox{\tiny{$p^2\to
0$}}}\kern0.5em}
\def\krto{ {\,\,\lower .8ex\hbox {$\longrightarrow \atop k \rightarrow 0$}\,\,}}

\def\Am#1#2#3{\widetilde A_{#1}^{#2}(#3)}

\def\bea{\begin{eqnarray}}
\def\beq{\begin{eqnarray}}

\def\eea{\end{eqnarray}}
\def\eeq{\end{eqnarray}}
\def\eq#1{eq.~(\ref{#1})}

\def\Eq#1{Eq.~(\ref{#1})}

\newcommand{\etabar}{\overline{\eta}}
\newcommand{\lwrsim}{\raise0.3ex\hbox{$<$\kern-0.75em\raise-1.1ex\hbox{$\sim$}}}
\newcommand{\lgrsim}{\raise0.3ex\hbox{$>$\kern-0.75em\raise-1.1ex\hbox{$\sim$}}}
\newcommand{\ghvertex}{\begin{picture}(100,25)(0,-3)
\SetWidth{1.2}
\DashArrowLine(12.5,0)(50,0){5}
\DashArrowLine(50,0)(87.5,0){5}
\Gluon(50,0)(50,25){-4}{3}
\CCirc(50,0){5}{Black}{Yellow}
\Text(12.5,-10)[l]{k}
\Text(87.5,-10)[r]{q}
\Text(60,20)[l]{q-k}
\end{picture}}
\newcommand{\ghost}{\begin{picture}(150,25)(0,0)
\SetWidth{1.2}
\DashArrowLine(12.5,0)(37.5,0){5}
\DashArrowLine(37.5,0)(112.5,0){5}
\DashArrowLine(112.5,0)(137.5,0){5}
\SetWidth{1}
\Vertex(37.5,0){2}
\Vertex(112.5,0){2}
\GlueArc(75,0)(37.5,0,90){-4}{6}
\GlueArc(75,0)(37.5,90,180){-4}{6}
\CCirc(75,37.5){10}{Black}{Blue}
\end{picture}}
\newcommand{\gluonTwoB}{\begin{picture}(150,25)(0,0)
\SetWidth{1.2}
\Gluon(12.5,0)(75,0){-4}{6}
\Gluon(75,0)(137.5,0){-4}{6}
\SetWidth{1}
\Vertex(75,0){3}
\GlueArc(75,20)(20,-90,90){4}{6}
\GlueArc(75,20)(20,90,270){4}{6}
\CCirc(75,40){10}{Black}{Blue}
\end{picture}}
\newcommand{\gluonTwoA}{\begin{picture}(150,25)(0,0)
\SetWidth{1.2}
\Gluon(12.5,0)(37.5,0){-4}{2}
\Gluon(37.5,0)(112.5,0){-4}{6}
\Gluon(112.5,0)(137.5,0){-4}{2}
\SetWidth{1}
\Vertex(37.5,0){2}
\Vertex(112.5,0){2}
\GlueArc(75,0)(37.5,0,90){-4}{6}
\GlueArc(75,0)(37.5,90,180){-4}{6}
\CCirc(75,37.5){10}{Black}{Blue}
\end{picture}}
\title{A Ghost Story II: Ghosts, Gluons and the Gluon condensate beyond the IR of QCD}
\ShortTitle{A Ghost Story II: Ghosts, Gluons and the Gluon condensate ...}
\author{\speaker{J.~Rodr\'iguez-Quintero}\\
  Dpto. F\'isica Aplicada, Huelva~\footnote{Fac. Ciencias Experimentales,
Universidad de Huelva, 21071 Huelva, Spain} \\
E-mail: \email{jose.rodriguez@dfaie.uhu.es}}
\author{Ph.~Boucaud, J.P.~Leroy, A.~Le~Yaouanc, J. Micheli, O.~P\`ene\\
LPT Orsay (CNRS)~\footnote{$^a$Laboratoire de Physique
 Th\'eorique, Unit\'e Mixte de Recherche 8627 du Centre National de 
la Recherche Scientifique 
Universit\'e de Paris XI, B\^atiment 210, 91405 Orsay Cedex,
France}\\
E-mail: \email{philippe.boucaud@th.u-psud.fr},
\email{leroy@th.u-psud.fr},\email{leyaouan@th.u-psud.fr},
\email{micheli@th.u-psud.fr},\email{olivier.pene@th.u-psud.fr}}
\author{F.~De Soto\\
U.P.O., Sevilla \\
E-mail: \email{fcsotobor@upo.es}}
\abstract{ Beyond the deep IR, the analysis of ghost and gluon propagators still 
keeps very interesting non-perturbative information. The Taylor-scheme coupling can 
be computed and applied to obtain the $\Lambda_{\rm QCD}$ parameter from Landau 
gauge lattice simulations. Furthermore, a dimension-two gluon condensate, that can 
be understood in the instanton liquid model, plays an important role in the game.
}
\FullConference{International Workshop on QCD Green's Functions, Confinement and Phenomenology\\
		 September 7-11 , 2009\\
ECT Trento, Italy}
\begin{document}

\section{Introduction}

Much work has been devoted in the last years to the study of the QCD running
coupling constant determined from lattice simulations, as well in its perturbative
regime~\cite{Bali:1992ru,Luscher:1993gh,deDivitiis:1994yp,Alles:1996ka,Boucaud:1
998bq,Boucaud:2000ey,OPEtree,OPEone,Sternbeck:2007br} as in the deep infrared domain~\cite{Boucaud:2002nc}.
Only very recently~\cite{Sternbeck:2007br,Boucaud:2008gn}, the Green's function approach to study the 
running coupling and then to estimate $\Lambda_{\overline{\rm MS}}$
has been pursued by exploiting a non-perturbative definition of the coupling 
derived from the ghost-gluon vertex. The very infrared domain for the running of the coupling so
defined has been discussed in the Olivier Pene's talk. We aim to deal here with the 
running behaviour of this ghost-gluon coupling beyond the IR domain, above roughly 2-3 Gev.

We will show that the analysis of {\it quenched} lattice simulations leads to a
non-perturbative determination of
the running coupling in terms of two-point ghost and gluon Green functions and 
to obtain $\Lambda_{\overline{\rm MS}}$ in pure Yang-Mills ($N_f=0$). Furthermore, a very realistic estimate
of $\Lambda_{\overline{\rm MS}}$, directly comparable with
experimental determinations, will become an immediate possibility thanks to the many unquenched
configurations which are presently available.%~\cite{Boucaud:2008xu}.

A precise determination of the non-perturbative coupling from the lattice 
also reveals a dimension-two non-zero gluon condensate in the landau gauge~\cite{Boucaud:2002nc}.
One needs then to describe the running with a formula including non-perturbative power corrections
to be confronted with lattice estimates of the coupling. This procedure constitutes an 
optimal method for the identification of $\Lambda_{\overline{\rm MS}}$ and of  
the gluon condensate~\cite{Boucaud:2008gn}. Much work has been also done to investigate its 
phenomenological implications in the gauge-invariant world~\cite{Gubarev:2000nz}. In particular, 
we will discuss the interpretation of this condensate in terms of the Yang-Mills semiclassical 
field background by applying the Instanton liquid model.

 \section{The ghost-gluon coupling}
 
 There is a large number of possibilities to define the QCD renormalized coupling constant, depending on the 
observable used to measure it and on the renormalization scheme. Actually, any observable which behaves, from 
the perturbative point of view, as $g$ provides a suitable definition for it. Among such quantities stand the 
3-gluon and the ghost-gluon vertices, which have been widely used by the lattice community to get a direct 
knowledge of $\alpha_s$ from  simulations. Of  course an important criterion to choose among those definitions 
will be how easy it is to connect it to other commonly used definitions, specially the $\overline{MS}$ one, 
and to extract from  it fundamental parameters like $\Lambda_{QCD}$.

A convenient class of renormalization schemes to work with on the lattice is made of the 
so-called $MOM$ schemes which are defined through the requirement that a given scalar 
coefficient function of the Green's function under consideration take  its tree-level 
value in a specific kinematical situation given  up to an overall ``renormalization scale''. 
To make the point clearer we recall 2 schemes which we have used in previous works on $\alpha_s$:
\begin{itemize}
\item The symmetric 3-gluon scheme in which one uses the 3-gluon vertex $\Gamma_{\mu\nu\rho}(p_1,p_2,p_3)$ 
with $p_1^2=p_2^2=p_3^2=\mu^2$
\item The asymmetric 3-gluon scheme ($\widetilde{MOM}$) in which  
the 3-gluon vertex $\Gamma_{\mu\nu\rho}(p_1,p_2,p_3)$ 
is used with $p_1^2=p_2^2=\mu^2,\,p_3^2=0$
\end{itemize}

In the present note we shall apply a specific $MOM$-type renormalization scheme defined 
by fixing the (ghost and gluon) propagators and the ghost-gluon vertex at the renormalization point.
Let us start by writing the ghost  and gluon propagators  in Landau gauge as follows,
\beq
\left( G^{(2)} \right)_{\mu \nu}^{a b}(p^2,\Lambda) &=& \frac{G(p^2,\Lambda)}{p^2} \ \delta_{a b} 
\left( \delta_{\mu \nu}-\frac{p_\mu p_\nu}{p^2} \right) \ ,
%\langle \widetilde{A^a_\mu}(-p) \widetilde{A^b_\nu}(p) \rangle \ ,
\nonumber \\
\left(F^{(2)} \right)^{a,b}(p^2,\Lambda) &=& - \delta_{a b} \ \frac{F(p^2,\Lambda)}{p^2} \ ;
\eeq
$\Lambda$ being some regularisation parameter ($a^{-1}(\beta)$ if, for instance, we specialise to lattice
regularisation). The renormalized dressing functions, $G_R$ and $F_R$ are defined through :
\beq\label{bar}
G_R(p^2,\mu^2)\ &= \ \lim_{\Lambda \to \infty} Z_3^{-1}(\mu^2,\Lambda) \ G(p^2,\Lambda)\nonumber\\
F_R(p^2,\mu^2)\ &= \ \lim_{\Lambda \to \infty} \widetilde{Z}_3^{-1}(\mu^2,\Lambda)\ F(p^2,\Lambda) \ ,
\eeq
with renormalization condition
\beq\label{bar2}
G_R(\mu^2,\mu^2)=F_R(\mu^2,\mu^2)=1 \ .
\eeq
Now, we will consider the ghost-gluon vertex which could be non-perturbatively obtained through 
a three-point Green function, defined by two ghost and one gluon fields, 
with amputated legs after dividing by two ghost and one gluon 
propagators. This vertex can be written quite generally as:

\beq\label{defGamma}
\widetilde{\Gamma}^{abc}_\nu(-q,k;q-k) = 
\ghvertex =
i g_0 f^{abc} 
\left( q_\nu H_1(q,k) + (q-k)_\nu H_2(q,k) \right) \ ,
\eeq

\noindent where $q$ is the outgoing ghost momentum and $k$ the incoming one, 
and renormalized according to:
\beq
\widetilde{\Gamma}_R=\widetilde{Z}_1 \Gamma.
\eeq
\noindent The vertex $\Gamma_\nu$  involves two independent scalar functions. In the MOM renormalization 
procedure $\widetilde{Z}_1$ is fully determined by demanding that one specific combination of those two form factors 
(chosen at one's will) be equal to its tree-level value for a specific kinematical 
configuration. 
We choose to apply MOM prescription for the scalar function $H_1+H_2$ that multiplies $q_\nu$ in \eq{defGamma} and 
the renormalization condition reads\footnote{In the case of zero-momentum gluon, an appropriate choice would be 
$\widetilde{Z}_1(\mu^2) H_1(q,q)|_{q^2=\mu^2}=1$. 
This would make the renormalized vertex equal to its tree-level value at 
the renormalization scale.} 
\beq\label{MOMT}
\left.(H^R_1(q,k) +  H^R_2(q,k))\right\vert_{q^2=\mu^2} = 
\lim_{\Lambda \to \infty}\widetilde{Z}_1(\mu^2,\Lambda)\left.(H_1(q,k;\Lambda) 
+  H_2(q,k;\Lambda))\right\vert_{q^2=\mu^2} =1,
\eeq
where we prescribe a kinematics for the subtraction point such that 
the outgoing ghost momentum is evaluated at the renormalization scale, while the incoming one, $k$, depends on the 
choice of several possible configurations; for instance: $k^2=(q-k)^2=\mu^2$ (symmetric configuration) or 
$k=0, \ (q-k)^2=\mu^2$ (asymmetric-ghost configuration). 
%The condition \eq{MOMT} is not suitable, for instance, for a MOM 
%renormalization of the ghost-gluon vertex at the following asymmetric substraction point: $q-k=0, \ k^2=q^2=\mu^2$.

On the other hand, the fields involved in the non-perturbative definition of the 
vertex $\Gamma_\nu$ in \eq{defGamma} can be directly renormalized by their renormalization constants, 
$Z_3$ and $\widetilde{Z}_3$, 
and the same MOM prescription applied to the scalar combination $H_1+H_2$ %in the r.h.s. of \eq{defGamma} 
also implies:
\beq\label{g2R}
g_R(\mu^2) &=& \lim_{\Lambda \to \infty} \ \widetilde{Z}_3(\mu^2,\Lambda) Z_3^{1/2}(\mu^2,\Lambda) g_0(\Lambda^2)   
\left. \left(  H_1(q,k;\Lambda) + H_2(q,k;\Lambda) 
\rule[0cm]{0cm}{0.5cm}  \right) \right|_{q^2 \equiv \mu^2} 
\nonumber \\
&=&  \ \lim_{\Lambda \to \infty} g_0(\Lambda^2) \ 
\frac{Z_3^{1/2}(\mu^2,\Lambda^2)\widetilde{Z}_3(\mu^2,\Lambda^2)}{ \widetilde{Z}_1(\mu^2,\Lambda^2)} \ .
\eeq
We combine both \eq{MOMT} and the first-line equation of (\ref{g2R}) to replace $H_1+H_2$ and obtain 
the second line that shows the well-known relationship $Z_g=(Z_3^{1/2} \widetilde{Z}_3)^{-1} \widetilde{Z}_1$, 
where $g_R=Z_g^{-1} g_0$.

We turn now to the specific $MOM$-type renormalization scheme defined by a {\bf{zero incoming ghost momentum}}. 
Since those kinematics are the ones (and the only ones) in which Taylor's well known non-renormalization theorem 
(cf. ref~\cite{Taylor}) is valid we shall refer to this scheme as to the $T$-scheme and the corresponding 
quantities will bear a $T$ subscript. Then, in  eq~(\ref{defGamma}), we set $k$ to $0$ and get 
\beq\label{defGammaT}
\widetilde{\Gamma}^{abc}_\nu(-q,0;q) = 
i g_0 f^{abc} 
\left(H_1(q,0) +  H_2(q,0) \right)\, q_\nu \ .
\eeq
Now, Taylor's theorem states that $H_1(q,0;\Lambda) +  H_2(q,0;\Lambda)$ is equal to 1 in full QCD for 
any value of $q$. Therefore, the renormalization condition \eq{MOMT} implies $\widetilde{Z}_1(\mu^2)=1 $ and then 
\beq\label{alpha} 
\alpha_T(\mu^2) \equiv \frac{g^2_T(\mu^2)}{4 \pi}=  \ \lim_{\Lambda \to \infty} 
\frac{g_0^2(\Lambda^2)}{4 \pi} G(\mu^2,\Lambda^2) F^{2}(\mu^2,\Lambda^2) \ ;
\eeq
where we also apply the renormalization condition for the propagators, eqs. (\ref{bar},\ref{bar2}), 
to replace the renormalization constants, $Z_3$ and $\widetilde{Z}_3$, by the bare dressing 
functions. The remarkable feature of  \eq{alpha} is that  it involves only $F$ and $G$ so that 
no measure of the ghost-gluon vertex is needed for the determination of the coupling constant.

Equation~(\ref{alpha}) has extensively been advocated and studied on the lattice 
(see for instance reference~\cite{vonSmekal:1997is}). However it must be stressed 
that the $T$-scheme is the {\bf only} one in which  $\widetilde{Z_1}=1$. 
Nevertheless the form (\ref{alpha}) is used quite often in this case (for a 
kinematical configuration other than T-scheme's) also as an approximation, 
specially in relation with the study of Dyson-Schwinger equations. 
An important remark is also in order here: in the very infrared domain, for 
phenomenological purposes (see for instance \cite{Binosi:2009qm}), the coupling can be more properly 
defined by pulling a massive gluon propagator out from the ghost-gluon Green function 
used to build it~\cite{Aguilar:2009nf}.

%%%%%%%%%%%%%%%%%%%%%%%%%%%%%%%%%%%%%%%%%%%%%%%%%%%%%%%%%%%%%%%%%%%%%%%%%%%%%%%%%%%%
\subsection{Pure perturbation theory}
\label{PTh}
%%%%%%%%%%%%%%%%%%%%%%%%%%%%%%%%%%%%%%%%%%%%%%%%%%%%%%%%%%%%%%%%%%%%%%%%%%%%%%%%%%%%

A standard four-loop formula describing the running for the $T$-scheme coupling,
%
%\begin{align}
\beq\label{betainvert}
%  \begin{split}
      \alpha_T(\mu^2) &=& \frac{4 \pi}{\beta_{0}t}
      \left(1 - \frac{\beta_{1}}{\beta_{0}^{2}}\frac{\log(t)}{t}
     + \frac{\beta_{1}^{2}}{\beta_{0}^{4}}
       \frac{1}{t^{2}}\left(\left(\log(t)-\frac{1}{2}\right)^{2}
     + \frac{\widetilde{\beta}_{2}\beta_{0}}{\beta_{1}^{2}}-\frac{5}{4}\right)\right) 
\nonumber \\
     &+& \frac{1}{(\beta_{0}t)^{4}}
 \left(\frac{\widetilde{\beta}_{3}}{2\beta_{0}}+
   \frac{1}{2}\left(\frac{\beta_{1}}{\beta_{0}}\right)^{3}
   \left(-2\log^{3}(t)+5\log^{2}(t)+
\left(4-6\frac{\widetilde{\beta}_{2}\beta_{0}}{\beta_{1}^{2}}\right)\log(t)-1\right)\right)
\nonumber \\
%     \end{split}
%\end{align}
%
% \beq
% h(t)=\frac 1 {\beta_0 t} \left( 1 - \frac{\beta_1}{\beta_0^2} \frac{\ln{t}} t 
% + \left( \frac {\beta_1}{\beta_0^2 t} \right)^2 \left( \left( \ln{t} - \frac 1 2 \right)^2 
% + \frac {\beta_2 \beta_0}{\beta_1^2} - \frac 5 4 \right) \right) \ ,
% \eeq
%
&\mbox{\rm{with}}& \ \ t=\ln{\frac{\mu^2}{\Lambda_T^2}} \ . 
%     \end{split}
%\end{align}
\eeq
is obtained by inverting the $\beta$-function of $\alpha_T$,
\beq\label{beta}
\beta_T(\alpha_T) \ = \ 
\frac{d\alpha_T}{d\ln{\mu^2}} \ = \ - 4 \pi \ 
\sum_{i=0} \widetilde{\beta}_i \left( \frac{\alpha_T} {4 \pi} \right)^{i+2} \ ;
\eeq
where, as explained in \cite{Boucaud:2008gn}, the coefficients $\widetilde{\beta}_i$ can 
be computed in terms of $\overline{\beta}_i$, those for the $\beta$-function of 
the coupling renormalizad according $\overline{\rm MS}$-scheme, $\overline{\alpha}$, and 
of the anomalous dimensions for gluon and ghost propagators,
\beq\label{betah}
\frac{1}{\alpha_T(\mu^2)} \ \frac{d\alpha_T(\mu^2)}{d\overline{\alpha}} & = & 
\frac{1}{\beta_{\overline{\rm MS}}(\overline{\alpha})} 
\left( 2 \ \lim_{\Lambda \to \infty} \frac d {d\ln{\mu^2}} \ln F(\mu^2,\Lambda)
+ \lim_{\Lambda \to \infty} \frac d {d\ln{\mu^2}} \ln G(\mu^2,\Lambda)
\right)
\nonumber 
\\
&=&  \frac{2 \widetilde{\gamma}(\overline{\alpha}) +\gamma(\overline{\alpha})}
{\beta_{\overline{\rm MS}}(\overline{\alpha})}
\rule[0.5cm]{0cm}{0.5cm} \ \ .
\eeq
Both anomalous dimensions need to be renormalized along MOM prescriptions 
({\it i.e.}, $G_R(\mu^2,\mu^2)=F_R(\mu^2,\mu^2)=1$) 
but expanded in terms of $\overline{\alpha}$. The coefficients so obtained  
(the details and results of the computation can be found in \cite{Boucaud:2008gn}) 
appear to agree with those directly obtained in ref.~\cite{Chetyrkin00} 
by the three-loop perturbative substracion of the ghost-gluon-gluon vertex in 
the QCD Lagrangian with the appropriate kinematical configuration ($T$-scheme).

%%%%%%%%%%%%%%%%%%%%%%%%%%%%%%%%%%%%%%%%%%%%%%%%%%%%%%%%%%%%%%%%%%%%%%%%%%%%
\subsection{OPE power corrections}
%%%%%%%%%%%%%%%%%%%%%%%%%%%%%%%%%%%%%%%%%%%%%%%%%%%%%%%%%%%%%%%%%%%%%%%%%%%%
\label{OPEsection}

In order to extend the description of the running coupling down to energies as low as 
possible (of the order of 3 GeV) and to take full advantage of  the lattice 
data we want to compare with, in order to  reduce the systematic uncertainties, 
it is mandatory to take into account  the gauge-dependent dimension-two OPE power 
corrections (cf.~\cite{OPEtree,OPEone,Boucaud:2002nc,Dudal:2002pq}) 
to  $\alpha_T$.

The leading power contribution to the ghost propagator, 
\beq\label{GhProp}
(F^{(2)})^{a b}(q^2) = \int d^4x e^{i q \cdot x} 
\langle \ T\left( c^a(x) \overline{c^b}(0) \right) \ \rangle 
\eeq
can be computed using the operator product 
expansion~\cite{Wilson69} (OPE), as is done in ref.~\cite{Boucaud:2005xn},
\beq\label{GhExp}
T\left( c^a(x)  \overline{c^b}(0) \right) = \sum_t \left(c_t\right)^{a b}(x) \ O_t(0);
\eeq
here $O_t$ is a local operator, regular when $x \to 0$, and  the Wilson coefficient $c_t$ 
contains the short-distance singularity. \Eq{GhExp} involves a full hierarchy of terms, ordered according to their mass-dimension, among which only  ${\bf 1}$ and $:A_\mu^a A_\nu^b:$ contribute to \eq{GhProp} in Landau gauge \footnote{The operators with 
an odd number of fields ($d=1,3/2$; $\partial_\mu A$ and $\partial_\mu \overline{c}$) cannot satisfy colour 
and Lorentz invariance and do not contribute  a non-zero non-perturbative expectation value, 
and  $\overline{c} c$ does not contribute either  because of the particular tensorial structure of the ghost-gluon 
vertex.} up to the order $1/q^4$.
Then, using \eq{GhExp} into \eq{GhProp}, we obtain:
\beq\label{OPE1}
(F^{(2)})^{a b}(q^2) &=& (c_0)^{a b}(q^2) \ + \ \left( c_2 \right)^{a b \sigma \tau}_{s t}(q^2)
\langle : A_\sigma^s(0) A_\tau^t(0): \rangle \ + \ \dots \nonumber \\ 
&=& (F_{\rm pert}^{(2)})^{a b}(q^2) \ + \ 
w^{a b} \ \frac{\langle A^2 \rangle}{4 (N_C^2-1)} \ + \ \dots 
\eeq
where 
\beq\label{OPE3}
w^{a b} \ &=& \ \left( c_2 \right)^{a b \sigma \tau}_{s t} \delta^{s t} g_{\sigma \tau} \ = \ 
\frac 1 2 \ \delta^{s t} g_{\sigma \tau} \frac{\int d^4x e^{i q \cdot x} \
\langle \Am{\tau'}{t'}{0} \ T\left( c^a \overline{c^b} \right) \ \Am{\sigma'}{s'}{0} \rangle_{\rm connected}}
{{G^{(2)}}_{\sigma \sigma'}^{s s'} {G^{(2)}}_{\tau \tau'}^{t t'} } \nonumber \\
&=&   2 \times \rule[0cm]{0cm}{1.7cm} \ghost,
\eeq
and the SVZ factorisation~\cite{SVZ} is invoked to compute the Wilson coefficients. 
Thus, one should compute the ``{\it sunset}'' diagram of the last line of \eq{OPE3}, that binds
the ghost propagator to the gluon condensate (where the blue bubble means contracting the 
color and lorentz indices of the incoming legs with $1/2 \delta_{st} \delta_{\sigma \tau}$) 
to obtain the leading non-perturbative contribution 
(of course, the first Wilson coefficient gives trivially the perturbative propagator).

Finally, after the renormalization of the $A^2$-condensate at the subtraction point $q^2=\mu^2$, 
according to the MOM scheme definition, the ghost dressing function is written as:
\beq\label{Z3fantome}
F_R(q^2,\mu^2) \ = \ F_{R, {\rm pert}}(q^2,\mu^2) \
\left(  1 + \frac{3}{q^2} \frac{g^2_R \langle A^2 \rangle_{R,\mu^2}} {4 (N_C^2-1)} \right) \ ,
\eeq
where the multiplicative correction to the purely perturbative $F_{R,{\rm pert}}$ is determined 
up to corrections of the order $1/q^4$ or $\ln{q/\mu}$.
As far as we do not deal with the anomalous dimension of the $A^2$ operator, 
factorising this purely perturbative ghost dressing function in \eq{Z3fantome} is 
a matter of choice. However, the Wilson coefficient is also computed 
at the leading logarithm in ref.~\cite{Boucaud:2008gn} and \eq{Z3fantome} appears then to 
be a very good approximation up to this order.

We can handle in the same way (see refs.\cite{OPEtree,OPEone})  
the OPE power correction to the gluon propagator, 
\beq\label{OPEgl1}
(G^{(2)}_R)_{\mu\nu}^{a b}(q^2,\mu^2)&=& (G_{R,\rm pert}^{(2)})_{\mu\nu}^{a b}(q^2,\mu^2) \ + \ 
\left(w_{\mu\nu}^{a b}\right)_{R,\mu^2} \ \frac{\langle A^2 \rangle_{R,\mu^2}}{4 (N_C^2-1)} \ + \ \dots 
\ ,
\eeq
and  obtain 
\beq\label{OPEgl3}
w_{\mu\nu}^{a b} \ &=&  \gluonTwoB
+ \
 2 \times \rule[0cm]{0cm}{1.7cm} \gluonTwoA \
\nonumber \\
&=&\rule[0.5cm]{0cm}{0.5cm} \frac{3 g^2}{q^2} \ (G_{\rm pert}^{(2)})_{\mu\nu}^{ab} \ .
\eeq
Then, after renormalization and appropriate projection, one gets for the gluon dressing function:
\beq\label{Z3glue}
G_R(q^2,\mu^2) \ = \ G_{R, {\rm pert}}(q^2,\mu^2) \
\left(  1 + \frac{3}{q^2} \frac{g^2_R \langle A^2 \rangle_{R,\mu^2}} {4 (N_C^2-1)} \right) \ .
\eeq

Finally, putting together the defining relation \eq{alpha} and the results eqs.~(\ref{Z3fantome},\ref{Z3glue}) we get
\beq\label{alphahNP}
\alpha_T(\mu^2) &=& \lim_{\Lambda \to \infty} 
\frac{g_0^2}{4 \pi} F^2(\mu^2,\Lambda) G(\mu^2,\Lambda) \nonumber \\
&=& 
\overbrace{\lim_{\Lambda \to \infty} 
\frac{g_0^2}{4 \pi} F^2(q_0^2,\Lambda) G(q_0^2,\Lambda)}^{\displaystyle
\alpha^{\rm pert}_T(q_0^2)} \ 
F^2_R(\mu^2,q_0^2) \ G_R(\mu^2,q_0^2)
\nonumber \\
&=&
\underbrace{
\alpha^{\rm pert}_T(q_0^2) F^2_{R,{\rm pert}}(\mu^2,q_0^2) \ G_{R,{\rm pert}}(\mu^2,q_0^2) 
}_{\displaystyle \alpha^{\rm pert}_T(\mu^2)}
\ 
\left( 
 1 + \frac{9}{\mu^2} \frac{g^2_T(q_0^2) \langle A^2 \rangle_{R,q_0^2}} {4 (N_C^2-1)}
\right) \ ,
\eeq
where $q_0^2 \gg \Lambda_{\rm QCD}$ is some perturbative scale and
the running of the perturbative part of the evolution, $\alpha_T^{\rm pert}$, 
is of course described by the \eq{betainvert} in the previous section.
Again, the Wilson coefficient at leading logarithm for the T-scheme MOM running coupling 
is obtained in \cite{Boucaud:2008gn} and found not to induce a significant effect, provided 
that the coupling multiplying $A^2$ inside the bracket is taken to be 
renormalized also in T-scheme. Thus, \eq{alphahNP} describes pretty well the running of $\alpha_T$ 
roughly above 3 Gev.

%%%%%%%%%%%%%%%%%%%%%%%%%%%%%%%%%%%%%%%%%%%%%%%%%%%%%%%%%%%%%%%%%%%%%%%%%%%%%
\section{Data Analysis by the ``{\it plateau}'' method}
%%%%%%%%%%%%%%%%%%%%%%%%%%%%%%%%%%%%%%%%%%%%%%%%%%%%%%%%%%%%%%%%%%%%%%%%%%%%%
\label{Dat-An}

In the following, as done in \cite{Boucaud:2008gn}, we will apply 
a ``{\it plateau}''-procedure exploiting \eq{alphahNP} to get a reliable estimate of the 
$\Lambda_{\rm QCD}$-parameter from lattice data. 
The goal being to get a trustworthy  estimate of the 
$\Lambda_{\overline{\rm MS}}$-parameter, one could attempt to do it by inverting 
the perturbative formula \eq{betainvert} and  using in the {\it inverted} formula
 the lattice estimates of the running coupling obtained by means of  \eq{alpha} for as 
many lattice momenta as possible. Then, one should look for 
a ``{\it plateau}'' of $\Lambda_{\overline{\rm MS}}$ in terms of 
momenta in the high-energy perturbative regime (this was done 
with the coupling defined by the three-gluon vertex in \cite{Alles:1996ka,Boucaud:1998bq}). 
In the next subsection, fig.~\ref{plot-plateau}.(a) shows the estimates of 
$\Lambda_{\overline{\rm MS}}$ so calculated for the lattice data presented 
in ref.~\cite{Boucaud:2005xn,Boucaud:2005gg} over $9 \ \lwrsim p^2 \ \lwrsim \ 33$~GeV$^2$.

However, in order to take advantage of the largest possible momenta window one can use instead 
\eq{alphahNP}. In this way we shall hopefully be able to extend towards {\it low} momenta the region 
over which to look for the best possible values of
the gluon condensate and of $\Lambda_{\overline{\rm MS}}$~\footnote{This increases the statistics and reduces errors. It also avoids 
some possible systematic deviation appearing when lattice momentum components, in lattice units, approach 
 $\pi/2$ (Brillouin's region border).}.
In other words, one requires the best-fit to a constant of 
\beq
(x_i,y_i) \equiv \left( p^2_i,\Lambda(\alpha_i) \right) \ , \ \ \ \ %\nonumber \\
{\rm with:} \ \ \ \ \ \alpha_i = \frac{\alpha_{\rm Latt}(p^2_i)}{\displaystyle 1+\frac{c}{p^2_i}} \ ;
\eeq
where $\Lambda(\alpha)$ is obtained by inverting the 
perturbative four-loop formula, \eq{betainvert}, 
and $c$ results from the best-fit (it appeared written in terms of the gluon condensate in 
\eq{alphahNP}~). Thus, $\Lambda(\alpha)$ reaches a ``{\it plateau}'' (if it does) behaving 
in terms of the momentum as a constant that we will take as our estimate of 
$\Lambda_{\overline{\rm MS}}$. Of course, this is nothing but a fitting strategy for a 
2-parameters ($\Lambda_{\overline{\rm MS}}$ and $\langle A^2 \rangle$) fit of the 
estimates of \eq{alpha} from lattice data.

%%%%%%%%%%%%%%%%%%%%%%%%%%%%%%%%%%%%%%%%%%%%%%%%%%%%%%%%%%%%%%%%%%%%%%%%%%%%
\subsection{Results for pure Yang-Mills ($N_f=0$)}
%%%%%%%%%%%%%%%%%%%%%%%%%%%%%%%%%%%%%%%%%%%%%%%%%%%%%%%%%%%%%%%%%%%%%%%%%%%%

The quenched lattice data that we will exploit now were presented for the first time in 
ref.~\cite{Boucaud:2005gg}. We refer to this work for all the details concerning the lattice 
implementation: algorithms, action, Faddeev-Popov operator inversion, etc.
The parameters of the whole set of simulations are described in 
table~\ref{setup} 

%%%%%%%%%%%%%%%%%%%%%%%%%%%%%%%%%%%%%%%%%%%%%%%%%%%%%%%%%%%%%%%%%%%%%%%%%%%%%
\begin{table}[ht]
\centering
\begin{tabular}{c|c||c|c}
\hline
$\beta$ & Volume & $a^{-1}$ (GeV) & Number of confs.
\\ \hline
$6.0$ &  $16^4$ & $1.96$ & $1000$
\\ \hline
$6.0$ &  $24^4$ & $1.96$ & $500$
\\ \hline
$6.2$ &  $24^4$ & $2.75$ & $500$
\\ \hline
$6.4$ &  $32^4$ & $3.66$ & $250$
\\ 
\hline
\end{tabular}
\caption{Run parameters of the exploited data.}
\label{setup}
\end{table}

\begin{figure}[ht]
\begin{center}
\begin{tabular}{cc}
\begin{tabular}{c}
\includegraphics[width=7cm]{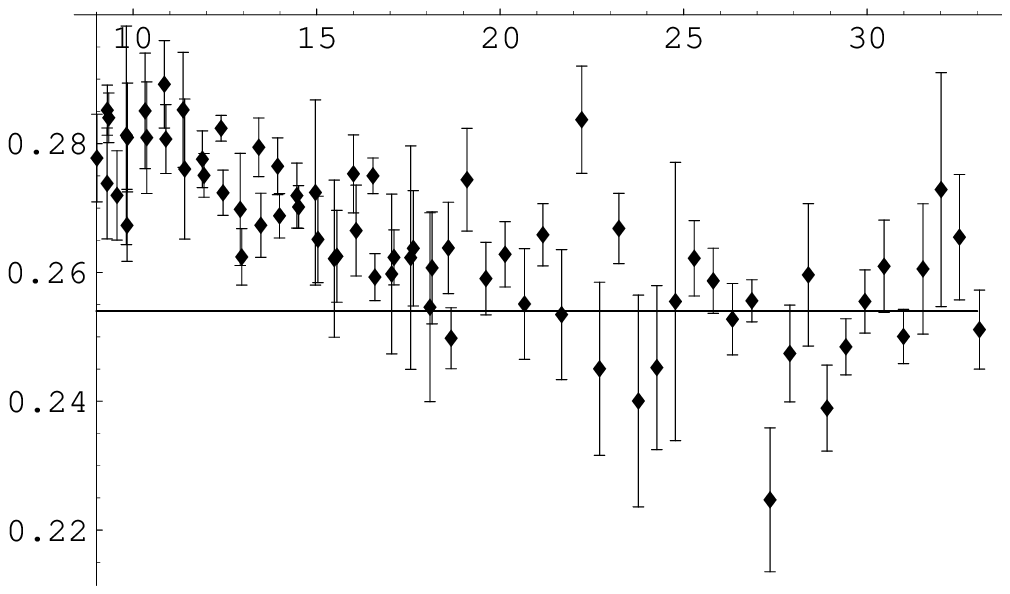}
\\  \rule[-1.9cm]{0cm}{3.8cm} (a) 
\end{tabular}
&
\begin{tabular}{c}
\includegraphics[width=7cm]{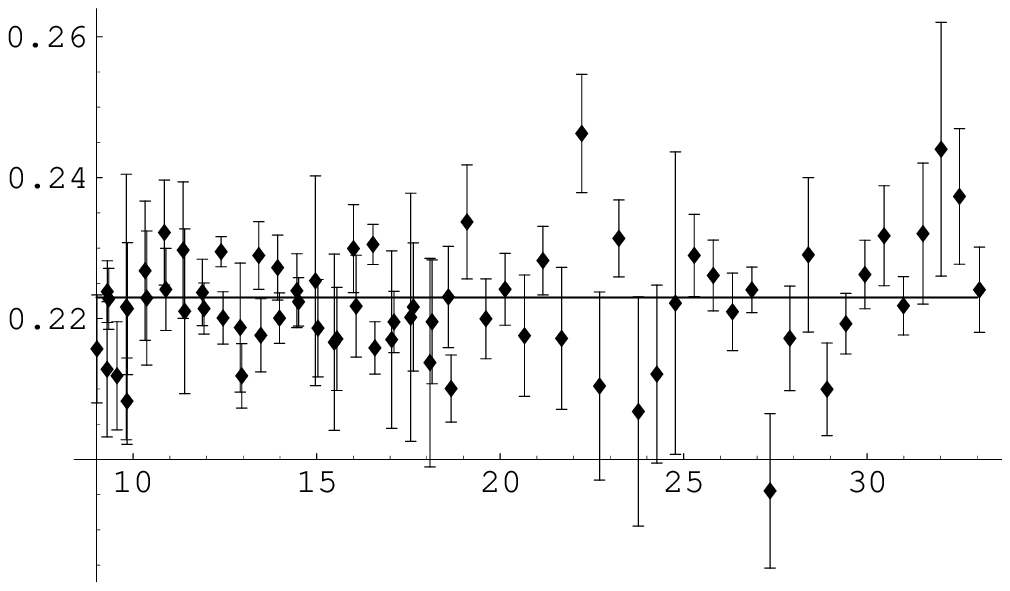}
\\ (b)
\end{tabular}
\end{tabular}
\vspace*{-2.1cm}
\end{center}
\caption{\small (a) Plot of  $\Lambda_{\overline{\rm MS}}$ (in GeV) computed by the inversion of 
the four-loop perturbative formula as a function  of the square of the momentum (in GeV$^2$); 
 the coupling is estimated from the lattice data through the perturbative formula obtained in the text.
(b) Same as plot (a) except for applying the non-perturbative formula obtained in the text for 
the coupling and looking for the gluon condensate generating the best plateau over 
$9 \ \lwrsim \ p^2 \ \lwrsim \ 33$ GeV$^2$.} 
\label{plot-plateau}
\end{figure}

In fig.~\ref{plot-plateau}.(a), we show  the estimates of $\Lambda_{\overline{\rm MS}}$ 
obtained when interpreting the lattice coupling computed by \eq{alpha} for any momentum 
$9 \ \lwrsim p^2 \ \lwrsim 33$ GeV$^2$ in terms of the  {\it inverted}  four-loop perturbative 
formula for the coupling, \eq{betainvert}. The estimates systematically decrease as 
the squared momentum increases until around 22 GeV$^2$;  above this value, only a 
noisy pattern results. In fig.~\ref{plot-plateau}.(b), the same is plotted but inverting  instead
the non-perturbative formula including power corrections, \eq{alphahNP}. The value of the gluon 
condensate has been determined by requiring a ``{\it plateau}'' to exist 
(as explained in the previous section) over the total momenta window.

%%%%%%%%%%%%%%%%%%%%%%%%%%%%%%%%%%%%%%%%%%%%%%%%%%%%%%%%%%%%%%%%%%%%%%%%%%%%%%%%%%%%%%%%%%
%%%%%%%%%%%%%%%%%%%%%%%%%%%%%%%%%%%%%%%%%%%%%%%%%%%%%%%%%%%%%%%%%%%%%%%%%%%%%%%%%%%%%%%%%%

\begin{figure}[htb]
\begin{center}
\begin{tabular}{cc}
\includegraphics[width=7cm]{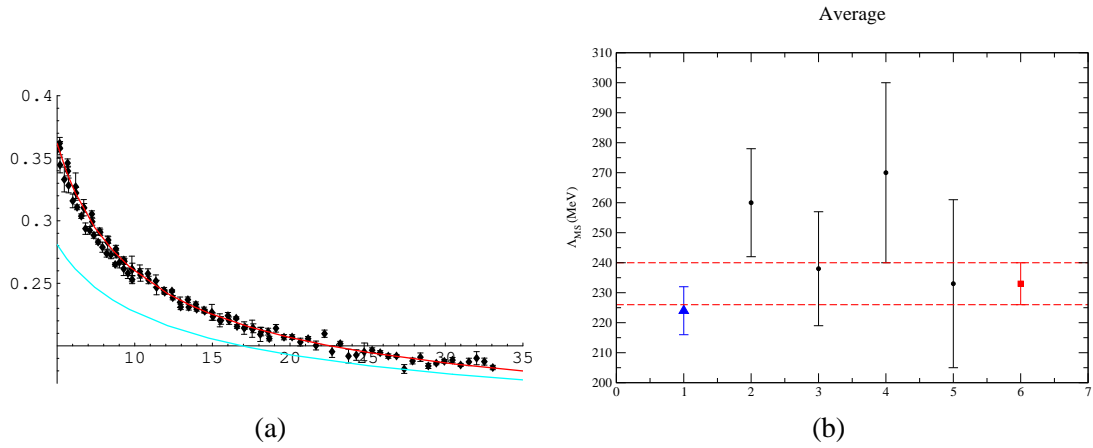} &
\includegraphics[width=7cm]{lambdas-wav.eps}
\\ 
(a) & (b)
\end{tabular}
\end{center}
\caption{\small (a) Plot of $\alpha_T$ in terms of the square of the 
renormalization momentum: the red solid line is computed with the non-perturbative formula
with  $\Lambda_{\overline{\rm MS}}=224$ MeV, the blue one with the perturbative one for the 
same $\Lambda_{\overline{\rm MS}}$ and the data are obtained from the lattice data 
set-up described in the text. (b)  Comparison with previous published estimates 
of $\Lambda_{\overline{\rm MS}}$ in pure Yang-Mills; the blue triangle stands for the estimate 
in this work and the red square for the {\it average} of the five estimates presented
in the plot. 
%The 1-$\sigma$ error interval for the average (dashed red line) were estimated 
%by treating the errors in tab.~\ref{comp} as purely statistical ones
}
\label{plot-alpha}
\end{figure}

One should realize that the non-perturbative 
analysis seems to indicate that the perturbative regime is far from being achieved 
at $p=5$ GeV.  This is also illustrated by figure~\ref{plot-alpha}.a in which, adopting 
for $\Lambda_{\overline{\rm MS}}$ the value $224$ MeV which results from the 
non-perturbative analysis, we plot against the square of the renormalization momentum 
the coupling constant as computed by means of the non-perturbative formula (\ref{alphahNP}) (red curve) 
and of the perturbative one  (\ref{betainvert}) (blue curve). Displayed are also the lattice data, 
{\it i.e.} the values of $\alpha_T$ obtained from \eq{alpha}. 

%%%%%%%%%%%%%%%%%%%%%%%%%%%%%%%%%%%%%%%%%%%%%%%%%%%%%%%%%%%%%%%%%%%%%%%%%%%%%%%%%%%%%%%%%%
%%%%%%%%%%%%%%%%%%%%%%%%%%%%%%%%%%%%%%%%%%%%%%%%%%%%%%%%%%%%%%%%%%%%%%%%%%%%%%%%%%%%%%%%%%

Thus,  one can conclude that our best-fit parameters incorporating only~\footnote{The error 
analysis is deeply discussed in \cite{Boucaud:2008gn}.} statistical errors are:
\beq\label{best-fit}
\Lambda_{\overline{\rm MS}}^{Nf=0 }&=&224^{+8}_{-5} \ \rm{MeV} \nonumber \\
g^2_T \langle A^2 \rangle_R &=& 5.1^{+0.7}_{-1.1} \ \rm{GeV}^2 \ .
\eeq
These values are in very good agreement with the previous estimates
from quenched lattice simulations of the three-gluon Green 
function~\cite{OPEtree,OPEone} or, in the case of $\Lambda_{\overline{\rm MS}}$, 
from the implementation of the Schr\"odinger functional method~\cite{Luscher:1993gh}, although 
slightly larger than the one obtained by the ratio of ghost and gluon 
dressing functions~\cite{Boucaud:2005xn} (see fig.~\ref{plot-alpha}.(b) and tab.~\ref{comp}). 
\begin{table}
\begin{center}
\begin{tabular}{|c||c|c|c|c|c|}
\hline
& $F^2G$ \cite{Boucaud:2008gn} & Asym. 3-g \cite{OPEone} & Sym. 3-g \cite{OPEone} &  $F/G$~\cite{Boucaud:2005xn} & \cite{Luscher:1993gh} \\ \hline
$\Lambda_{\overline{\rm MS}}$ (MeV) & 224$^{+8}_{-5}$ &  260(18) & 233(28) & 270(30) & 238(19) \\
\hline
$\sqrt{\langle A^2 \rangle_{R,\mu}}$ (GeV) & 1.64(17) & 2.3(6) & 1.9(3) & 1.3(4) & -- \\
\hline
\end{tabular}
\end{center}
\caption{\small Comparison of the estimate of $\Lambda_{\overline{\rm MS}}$ obtained from the 
analysis of the ghost-gluon vertex (first column) and others from literature. The renormalization 
point is $\mu=10$ GeV.
%the asymmetric 3-gluon vertex (second), 
%the symmetric 3-gluon vertex 
%(third), the ratio of gluon and ghost dressing functions (fourth) and 
%with the Schr\"odinger functional method (last). The gluon condensate $\langle A^2 \rangle_{R,\mu}$ has 
%been obtained at the renormalization momentum $\mu=10$ GeV
%, for the sake of comparison with the other 
%estimates, 
%from the best-fit estimate by applying $g^2(\mu^2=100$ GeV$^2)/4 \pi=0.15$.
}
\label{comp}
\end{table}

\section{About the nature and the size of the gluon condensate}

\begin{figure}[htb]
\begin{center}
\begin{tabular}{ccc}
\includegraphics[width=3.75cm]{propag2458.eps} &
\includegraphics[width=3.75cm]{vertex2458.eps} &
\includegraphics[width=6cm]{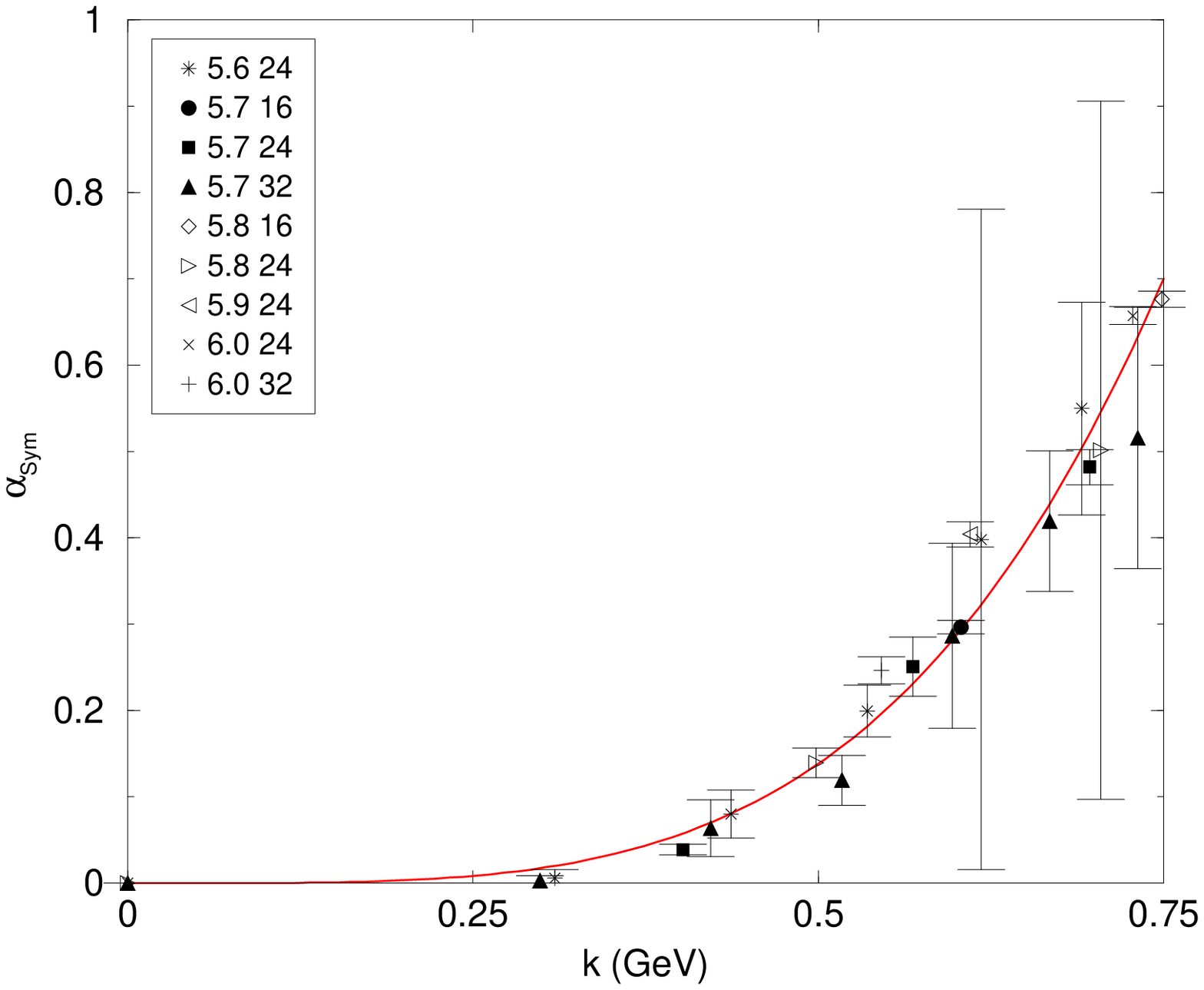}
\\ 
& (a) \rule[0cm]{1.5cm}{0cm}& (b)
\end{tabular}
\end{center}
\caption{\small (a) two and three-gluon Green functions after cooling: they reach their expected 
power-law when the number of cooling sweeps increases. (b) The three-gluon coupling defined 
in the text: it follows the expected low-momentum $k^4$ power-law with $n=5.27(4)$ fm$^{-4}$. 
%The 1-$\sigma$ error interval for the average (dashed red line) were estimated 
%by treating the errors in tab.~\ref{comp} as purely statistical ones
}
\label{plot-3g}
\end{figure}

The nature of the dimension-two gluon condensate, as well as its possible phenomenological 
implications, have been discussed in many works in the last few years (see for instance 
\cite{Gubarev:2000nz,Dudal:2002pq}). In particular, we presented some indications supporting 
the idea that the low-momentum gluon correlation functions could be nicely described in terms of the 
semiclassical instanton background for the gauge field~\cite{Boucaud:2004zr}, and used an instanton liquid 
picture to estimate the size for this gluon condensate in Yang-Mills~\cite{Boucaud:2002nc}.
Indeed, the gauge field in the instanton picture and within the sum-ansatz approach, 
can be written in the singular Landau gauge as
\beq
g A_\mu^a= 2 \sum_i R^{a \alpha}_{(i)} \etabar_{\mu\nu}^\alpha \frac{ \left(x^\nu-z_\nu^i \right)}{|x-z^i|^2} 
\phi\left(\frac{|x-z^i|}{\rho_i} \right) \  ,
\label{Bsu2}
\eeq
where $g=(6/\beta)^{1/2}$ is the bare gauge coupling in terms of the lattice parameter $\beta$,
$\etabar$ is known as 't Hooft  symbol and $R^{a \alpha}$ represents the color rotations
embedding the canonical SU(2) instanton solution in the SU(3) gauge  group, $\alpha=1,\cdots,3$
($a=1,\cdots,8$) being an SU(2) (SU(3)) color index. The sum is extended over all the instantons
and  anti-instantons (we should then replace the 't Hooft symbol $\etabar$ by $\eta$) in the
classical background of the  gauge configuration. $\phi(x)$ is the instanton profile function.
If we consider the profile of the BPST solution for an isolated instanton, we get
\beq 
g^2<A^2> \ \equiv \ \frac{N_I+N_A}{V} \ 
\int d^4x \sum_{\mu,a} g A^{a}_\mu \ g A^{a}_\mu
\ = \ 12 \pi^2 \rho^2 \frac{N_I+N_A}{V} \ = \ 12 \pi^2 \rho^2 n \ ;
\label{E9}
\eeq
where $N_I$ ($N_A$) stands for the total number of instantons (anti-instantons). On the other hand, 
if we neglect instanton position and color correlations, eq. (\ref{Bsu2}) leads for the $m$-gluon Green 
function to 
\beq
G^{(m)}(k^2) &=&  n \frac{4 k^2}{m} \left( \frac{\beta}{96 k^2} \right)^{m/2}  
<\rho^{3m}  I(k \rho)^m > \ , \nonumber \\
\rm{where} \ \ I(s) &=&  \frac{8 \pi^2}{s} \int_0^\infty\ z dz  J_2(sz) \, \phi(z) \ ,
\label{green}
\eeq
for $m=2,3$; $n$ being the instanton density. It depends on
the functional $I(k\rho)$ of the general instanton profile, $\phi(x)$, 
and $< \ \cdots \ >$ means the average over instanton sizes with a given normalised instanton radius 
distribution, $\mu(\rho)$. Then, two interesting limits appear where some results not depending 
on the instanton profile can be obtained:
\begin{itemize}
\item For a sharp radius distribution, the particular combination of two and three-gluon Green functions 
defining the three-gluon running coupling in ref.~\cite{Boucaud:1998bq} gives~\cite{Boucaud:2002fx}
\beq\label{3g}
\alpha_{\rm 3g}(k^2) = \frac{k^6}{4 \pi} \frac{\left(G^{(3)}\right)^2}{\left(G^{(2)}\right)^3} 
\ = \ \frac{k^4}{18 \pi n} \ ;
\eeq
\item For $k \rho \gg 1$, as $I(s)$ asymptotically behaves as $16\pi^2/s^3$ in the large $s$ limit, 
one obtains
\beq
\label{Gmfin}
G^{(m)}(k^2) \ \simeq \  n \frac 4 m \left(\frac{8 \beta}{3} \right)^{m/2} k^{2-4m} \ .
\eeq
\end{itemize}
Thus, \eq{Gmfin} provides us with large-momentum limits for the two and three-gluon Green functions 
behaviour which do not depend on the radius distribution nor on the instanton profile. However, 
the large-momentum lattice correlation function being dominated by the short-distance quantum 
fluctuations, whether such a behaviour occurs can be only detected after performing some ``cooling'' 
procedure~\cite{Teper:1985rb} to kill the higher energy modes. 
This is done in \cite{Boucaud:2004zr} and, as can be seen 
in fig.~\ref{plot-3g}.a, the expected $k^{-6}$ ($k^{-10}$) power behaviour clearly emerges for the 
two-gluon (three-gluon) Green function after ``cooling''. This is a good indication for the success of 
the instanton picture in describing the gluon correlation functions. However, as the ``cooling'' 
has been proved to alter the configuration (instanton sizes become distorted, instanton and anti-instanton 
anhiliate to each other...), the power-law given by \eq{3g}, which is thought to be followed by 
the ``uncooled'' gluon correlators in the low-momentum regime, offers a more reliable 
``instanton detector''. In ref.~\cite{Boucaud:2002fx}, \eq{3g} is shown to work for a three-gluon coupling 
computed from several lattice simulations (see fig.~\ref{plot-3g}.b taken 
from \cite{Boucaud:2002fx}) and the instanton density is estimated 
to be $n \simeq 5$ fm$^{-4}$.

%%%%%%%%%%%%%%%%%%%%%%%%%%%%%%%%%%%%%%%%%%%%%%%%%%%%%%%%%%%%%%%%%%%%%%%%%%%%%%%%%%%%%%%%%%
%%%%%%%%%%%%%%%%%%%%%%%%%%%%%%%%%%%%%%%%%%%%%%%%%%%%%%%%%%%%%%%%%%%%%%%%%%%%%%%%%%%%%%%%%%

Then, this estimate of the instanton density and the average instanton radius, 
$\overline{\rho}\simeq 0.4$ fm (measured, for instance, in \cite{Boucaud:2004zr} and being close to 
the phenomenological prediction, $\simeq 1/3$ fm), can be applied to \eq{E9} to 
give: $g^2 \langle A^2 \rangle \simeq 4 {\rm GeV}^2$.  There is of course
no exact recipe to compare this estimate with the OPE one, since the separation
between the semiclassical non perturbative domain and the perturbative one cannot be
exact~\footnote{One may appeal to the fact that at the
renormalisation point $\mu$, the radiative
corrections are minimised; therefore
a semiclassical estimate must best
correspond to $\langle A^2 \rangle_{\rm R,\mu}$ at some reasonable
$\mu$, which one could guess to be a typical scale of the problem 
as $1/\rho$ or some gluon mass.}. 
However, both lie prettily on the same ballpark.

\section{Conclusion}

We have demonstrated that, in the particular $T$-scheme, the coupling defined from 
the ghost-gluon vertex is obtained by only dealing with two-point Green functions. 
Some interesting non-perturbative information can be furthermore distilled from the 
running analysis of this coupling beyond the deep IR: the $\Lambda_{\rm QCD}$ 
parameter (usually expressed in the $\overline{\rm MS}$-scheme), computed here 
for pure Yang-Mills from Landau gauge lattice simulations, and a gauge-dependent 
dimension-two gluon condensate. The latter is interpreted and sized by invoking an 
instanton liquid picture, which successfully describes the low-momentum gluon correlations.

  \end{document}